\documentclass{article}
\usepackage{epsfig}
\usepackage{amssymb}
\catcode`\@=11
\def\@normalsize{\@setsize\normalsize{15pt}\xiipt\@xiipt
\abovedisplayskip 14pt plus3pt minus3pt%
\belowdisplayskip \abovedisplayskip
\abovedisplayshortskip  \z@ plus3pt%
\belowdisplayshortskip  7pt plus3.5pt minus0pt}
\def\small{\@setsize\small{13.6pt}\xipt\@xipt
\abovedisplayskip 13pt plus3pt minus3pt%
\belowdisplayskip \abovedisplayskip
\abovedisplayshortskip  \z@ plus3pt%
\belowdisplayshortskip  7pt plus3.5pt minus0pt
\def\@listi{\parsep 4.5pt plus 2pt minus 1pt
            \itemsep \parsep
            \topsep 9pt plus 3pt minus 3pt}}
\def\underline#1{\relax\ifmmode\@@underline#1\else
        $\@@underline{\hbox{#1}}$\relax\fi}
\@twosidetrue \relax \catcode`@=12
\catcode`\@=11

\def\section{\@startsection{section}{1}{\z@}{3.5ex plus 1ex minus
   .2ex}{2.3ex plus .2ex}{\large\bf}}

\def\ps@headings{\def\@oddfoot{}\def\@evenfoot{}
\def\@oddhead{\hbox{}\hfill
        \makebox[.5\textwidth]{\raggedright\ignorespaces --\thepage{}--
        \hfill }}
\def\@evenhead{\@oddhead}
\def\subsectionmark##1{\markboth{##1}{}}
} \ps@headings \catcode`\@=12 \relax

\def\figcap{\section*{Figure Captions\markboth
        {FIGURECAPTIONS}{FIGURECAPTIONS}}\list
        {Fig. \arabic{enumi}:\hfill}{\settowidth\labelwidth{Fig. 999:}
        \leftmargin\labelwidth
        \advance\leftmargin\labelsep\usecounter{enumi}}}
 \relax
\def\tablecap{\section*{Table Captions\markboth
        {TABLECAPTIONS}{TABLECAPTIONS}}\list
        {Table \arabic{enumi}:\hfill}{\settowidth\labelwidth{Table 999:}
        \leftmargin\labelwidth
        \advance\leftmargin\labelsep\usecounter{enumi}}}
 \relax
\def\reflist{\section*{References\markboth
        {REFLIST}{REFLIST}}\list
        {[\arabic{enumi}]\hfill}{\settowidth\labelwidth{[999]}
        \leftmargin\labelwidth
        \advance\leftmargin\labelsep\usecounter{enumi}}}
 \relax
\catcode`\@=11
\def\marginnote#1{}
\newcount\hour
\newcount\minute
\newtoks\amorpm
\hour=\time\divide\hour by60 \minute=\time{\multiply\hour by60
\global\advance\minute by- \hour}
\edef\standardtime{{\ifnum\hour<12 \global\amorpm={am}%
    \else\global\amorpm={pm}\advance\hour by-12 \fi
    \ifnum\hour=0 \hour=12 \fi
    \number\hour:\ifnum\minute<100\fi\number\minute\the\amorpm}}
\edef\militarytime{\number\hour:\ifnum\minute<100\fi\number\minute}

\def\draftlabel#1{{\@bsphack\if@filesw {\let\thepage\relax
  \xdef\@gtempa{\write\@auxout{\string
    \newlabel{#1}{{\@currentlabel}{\thepage}}}}}\@gtempa
    \if@nobreak \ifvmode\nobreak\fi\fi\fi\@esphack}
     \gdef\@eqnlabel{#1}}
\def\@eqnlabel{}
\def\@vacuum{}
\def\draftmarginnote#1{\marginpar{\raggedright\scriptsize\tt#1}}
\def\draft{\oddsidemargin -.5truein
        \def\@oddfoot{\sl preliminary draft \hfil
        \rm\thepage\hfil\sl\today\quad\militarytime}
        \let\@evenfoot\@oddfoot \overfullrule 3pt
        \let\label=\draftlabel
        \let\marginnote=\draftmarginnote
\def\@eqnnum{(\theequation)\rlap{\kern\marginparsep\tt\@eqnlabel}%
\global\let\@eqnlabel\@vacuum}  }
\def\preprint{\twocolumn\sloppy\flushbottom\parindent 1em
        \leftmargini 2em\leftmarginv .5em\leftmarginvi .5em
        \oddsidemargin -.5in    \evensidemargin -.5in
        \columnsep 15mm \footheight 0pt
        \textwidth 250mmin      \topmargin  -.4in
        \headheight 12pt \topskip .4in
        \textheight 175mm
        \footskip 0pt
\def\@oddhead{\thepage\hfil\addtocounter{page}{1}\thepage}
        \let\@evenhead\@oddhead \def\@oddfoot{} \def\@evenfoot{}
}
\def\titlepage{\@restonecolfalse\if@twocolumn\@restonecoltrue\onecolumn
     \else \newpage \fi \thispagestyle{empty}\c@page\z@
        \def\thefootnote{\fnsymbol{footnote}} }
\def\endtitlepage{\if@restonecol\twocolumn \else  \fi
        \def\thefootnote{\arabic{footnote}}
        \setcounter{footnote}{0}}  
\catcode`@=12 \relax

\def\ps@headings{\def\@oddfoot{}\def\@evenfoot{}
\def\@oddhead{\hbox{}\hfill
        \makebox[.5\textwidth]{\raggedright\ignorespaces --\thepage{}--
        \hfill }}
\def\@evenhead{\@oddhead}
\def\subsectionmark##1{\markboth{##1}{}}
} \ps@headings \relax
\newcommand{\newc}{\newcommand}

\newc{\ra}{\rightarrow}
\newc{\lra}{\leftrightarrow}
\newc{\beq}{\begin{equation}}
\newc{\be}{\begin{equation}}
\newc{\eeq}{\end{equation}}
\newc{\ee}{\end{equation}}
\newc{\bea}{\begin{eqnarray}}
\newc{\eea}{\end{eqnarray}}

\newc{\ome}{\omega}
\newc{\ba}{\begin{eqnarray}}
 \newc{\ea}{\end{eqnarray}}

\begin{document}
\def\firstpage#1#2#3#4#5#6{
\begin{titlepage}
\nopagebreak
\title{\begin{flushright}
        \vspace*{-0.8in}
{ \normalsize  
}
\end{flushright}
\vfill {#3}}
\author{\large #4 \\[1.0cm] #5}
\maketitle \vskip -7mm \nopagebreak
\begin{abstract}
{\noindent #6}
\end{abstract}
\vfill
\begin{flushleft}
\rule{16.1cm}{0.2mm}\\[-3mm]

\end{flushleft}
\thispagestyle{empty}
\end{titlepage}}

\def\simlt{\stackrel{<}{{}_\sim}}
\def\simgt{\stackrel{>}{{}_\sim}}
\date{}
\firstpage{3118}{IC/95/34} {\large\bf A D-brane inspired $U(3)_{C}\times
U(3)_{L}\times U(3)_{R}$ model}
 {G.K. Leontaris, J. Rizos}
{\normalsize\sl Theoretical Physics Division, Ioannina University,
GR-45110 Ioannina, Greece\\
\\ [2.5mm]
 }
{Motivated by D-brane scenarios, we consider a non-supersymmetric
model based on the gauge symmetry $U(3)_C\times U(3)_L\times
U(3)_R$ which is equivalent to the $SU(3)^3$ ``trinification''
model supplemented by three $U(1)$s. Two $U(1)$ combinations  are
anomalous while the third $U(1)_{{\cal Z}'}$ is anomaly free and
contributes to the hypercharge generator. This hypercharge
embedding correspods to $\sin^2\theta_W=\frac{6}{19}$ in the case
of full gauge coupling unification. The ${U(3)}^3$ symmetry is
broken down to the Standard Model by vev's of  two $(1,3,\bar
3)$-scalar multiplets supplemented by two Higgs fields in
$(1,3,1)$ and $(1,1,3)$ representations. The latter  break
$U(1)_{{\cal Z}'}$ and provide heavy masses to the extra lepton
doublets.  Fermions belong to $(3,\bar 3,1)+(\bar 3,1,3)+
(1,3,\bar 3)$ representations as in the trinification model.  The
model predicts a natural quark-lepton hierarchy, since quark
masses are obtained from tree-level couplings, while charged
leptons receive masses from fourth order Yukawa terms, as a
consequence of the extra abelian symmetries. Light Majorana
neutrino masses are obtained through a see-saw type mechanism
operative at the $SU(3)_R$ breaking scale of the order $M_R\ge
10^{9}$GeV.}

\vskip 3truecm

\section{Introduction}

 Extended objects of the non-perturbative sector of string theory, the so-called
  D-branes\cite{Polchinski:1996na}, appear to be a promising framework for model
 building. Intersecting D-branes in particular, can provide chiral fermions
 and gauge symmetries which contain the Standard Model spectrum  and the
 $SU(3)\times SU(2)\times U(1)$ symmetry as a subgroup and thus D-brane models
 appear to be natural candidates for phenomenological explorations. During the last
 years, particular supersymmetric or non-supersymmetric models have been
proposed~\cite{Antoniadis:2002en}-\cite{Leontaris:2001hh}, based on
various D-brane configurations, which exhibit a number of
interesting properties. A remarkable feature is that a low
unification scale can be possible, since the four-dimensional
gauge couplings depend on the volume of extra dimensions. In this case,
one can solve the hierarchy problem without supersymmetry.
Further, there exist anomalous $U(1)$ symmetries whose anomalies
are cancelled by a generalized Green-Schwarz mechanism; a linear
combination of these $U(1)$'s remains anomaly free and plays a
significant role in particular phenomenological explorations.

In the present work, we propose a model based on the gauge
symmetry ${U(3)}^3$ which can arise in a D-brane construction.
This symmetry contains as a subgroup the $SU(3)_C\times SU(3)_L
\times SU(3)_R$ symmetry, (trinification model) which has been
proposed long time ago~\cite{Glashow:1984gc,Rizov:1981dp} and
subsequently explored in a non-supersymmetric~\cite{Babu:1986gi},
or supersymmetric~\cite{Dvali:1994vj}-\cite{Maekawa:2002qv}
context. It has also been explored as a subgroup of the $E_6$
symmetry in field theory\cite{Gursey} or in the context of
strings~\cite{Greene:1986bm,Willenbrock:2003ca}. In this paper,
 we restrict to  the non-supersymmetric case, as the problem of
 supersymmetry breaking may be resolved at the D-brane
 level~\cite{Antoniadis:1999xk}.

In the D-brane analogue of the trinification model all fermions
are accommodated in the $(3,\bar 3,1)+(\bar 3,1,3)+(1,3,\bar 3)$
representations charged under three additional anomalous $U(1)$'s.
One linear combination of these $U(1)$ symmetries -hereafter
 $U(1)_{\cal Z'}$- is anomaly free and can serve as a
hypercharge component, leading to very interesting
phenomenological implications. Two Higgs fields ${\cal H}_a,
(a=1,2)$ in the representation $(1,3,\bar 3)$ (which is the same
one accommodating the lepton fields) are needed to break the
original symmetry down to the SM. In the D-brane construction, in
addition, two Higgs fields ${\cal H}_{\cal L}=(1,3,1)$ and ${\cal
H}_{\cal R}=(1,1,3)$ may also appear in the spectrum. When these
Higgs fields obtain vevs, they break $U(1)_{\cal Z'}$ and at the
same time provide heavy masses to a pair of the extra lepton
doublets.

Quark masses arise from tree-level couplings of the Yukawa
potential. The same coupling supports with a heavy mass an extra
color triplet. Due to the additional $U(1)$ symmetries, Yukawa
couplings for leptons are not allowed at tree-level, however, they
arise already at fourth order giving thus a natural explanation to
quark-lepton hierarchy. Further, higher order invariants for
Yukawa mass terms appear at even powers of the expansion parameter
$\frac{\langle {\cal H}\,{\cal H}^\dagger\rangle}{M_S^2}\le
10^{-1}$, (where $M_S$ is the string scale) ensuring thus the
validity of the perturbation theory in this model. The $U(3)^3$
model retains also all the interesting features of the
trinification model. Among them, the Higgs doublets and colored
fields are in different representations, therefore, no
doublet-triplet splitting is required.

The paper is organized as follows. In section 2 we give a
description of the $U(3)^3$ model motivated by D-brane scenarios
and present the fermion and Higgs spectrum. We
discuss the mixed anomaly cancellation and we identify the
anomaly-free $U(1)$ combination which contributes to the
hypercharge generator. We further discuss the gauge coupling
evolution and determine the range of the string scale as well as
the $SU(3)_L\times SU(3)_R$ intermediate breaking scale. In
section 3 we calculate the Yukawa potential and show that a
quark-lepton hierarchy arises, while all extra colored triplets
and doublets become massive at a high scale. We also discuss the
implications of the model for the neutrino masses. In section 4 we
present our conclusions.

\section{Description of the Model}

The model proposed here can be considered as a D-brane analogue of
the trinification ${SU(3)}^3$ model proposed in \cite{Glashow:1984gc,
Rizov:1981dp}. The minimal gauge symmetry obtained from D-branes which
has as a subgroup the $SU(3)^3$ model  is  $U(3)^3$.
Without going into details, we give here a brief description how
such a symmetry could arise in the context as a D-brane construction.
The basic ingredient is the brane stack, i.e., a certain number of
parallel, almost coincident D-branes.  A single D-brane carries a
$U(1)$ gauge symmetry which is the result of the reduction of the
ten-dimensional Yang-Mills theory. A stack of $N$ parallel branes
gives rise to a $U(N)$ gauge group.

We thus consider three stacks of D-branes, each stack containing 3
parallel almost coincident branes giving rise to the gauge symmetry
$$ U(3)_C\times U(3)_L\times U(3)_R\,.$$
The first $U(3)$ is related to  $SU(3)$ color, the second involves
the weak $SU(2)_L$ and the third is related to a possible
intermediate $SU(2)_R$ gauge group. Since $U(3)$ is equivalent to
$ SU(3)\times U(1)$ our D-brane construction  contains also three
extra $U(1)$ abelian symmetries. The $U(1)_C$ symmetry obtained
from the color $U(3)_C$ is related to the baryon
number~\cite{Antoniadis:2002en} which survives at low energies as
a global symmetry. There are two additional  abelian factors
originating from $U(3)_L$, $U(3)_R$ so the $U(3)^3$ symmetry can
be equivalently written
\ba
 SU(3)_C\times SU(3)_L\times SU(3)_R\times U(1)_C\times
U(1)_L\times U(1)_R\label{333111}
\ea
 \begin{figure}[h]
\centering
\includegraphics[scale=.9]{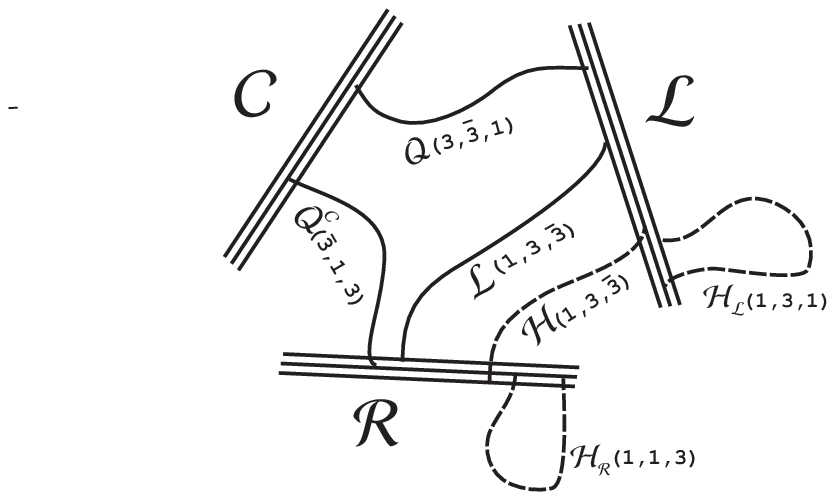}
\caption{Schematic representation of a $U(3)_C\times U(3)_L\times U(3)_R$  D-brane
configuration and the matter fields of the model.}
 \label{u333}
\end{figure}
In the D-brane context, matter fields appear as
 open strings having both their ends attached to some
of the brane stacks. For example, strings with both ends attached
on two different 3-brane stacks  belong to the $(3,\bar 3)$
multiplets of the corresponding gauge group factors.
The possible representations which arise in this scenario should
be appropriate to accommodate the standard model particles and Higgs fields.
As such candidates we choose the open strings that appear in figure \ref{u333}.
Under the decomposition (\ref{333111}) these lead to the following matter
representations
\ba
{\cal Q^{\hphantom{c}}}&=&(3,\bar 3,1)_{(+1,-1,\hphantom{+}0)}\label{QL}\\
{\cal Q}^c&=&(\bar 3,1,3)_{(-1,\hphantom{+}0,+1)}\label{QR}\\
{\cal L}^{\hphantom{c}}&=&(1,3,\bar
3)_{(\hphantom{+}0,+1,-1)}\label{LH}
\ea
We adopt a notation where the three first numbers refer to the
color, left and right $SU(3)$ gauge groups, while the three
indices correspond to the three $U(1)_{C,L,R}$ symmetries
respectively. It turns out that these three representations are
sufficient to accommodate all the fermions of the Standard Model.
In particular, the  representation (\ref{QL}) includes the quark
left handed doublets and an additional colored triplet with
quantum numbers as those of the down quark, while  representation
(\ref{QR}) contains the right-handed partners of (\ref{QL}).
Finally (\ref{LH})
 involves the lepton doublet, the right-handed electron and
its corresponding neutrino, two additional $SU(2)_L$ doublets and
another neutral state, called neutreto\cite{Glashow:1984gc}. For a
single family, we write the following assignment
\ba
(3,\bar 3,1)\,=\,\left(\begin{array}{ccc}
u_r&d_r&g_r\\
u_g&d_g&g_g\\
u_b&d_b&g_b
\end{array}\right),\;
(\bar 3,1,3)\,=\,\left(\begin{array}{ccc}
u^c_r&u^c_g&u^c_b\\
d^c_r&d^c_g&d^c_b\\
g^c_r&g^c_g&g^c_b
\end{array}\right),\;
(1,3,\bar 3)\,=\,\left(\begin{array}{ccc}
E^{c0}&E^-&e\\
E^{c+}&E^0&\nu\\
e^c&\nu^{c+}&\nu^{c-}
\end{array}\right).\label{familas}
\ea
In addition, (\ref{LH}) may also accommodate the Higgs
multiplets responsible for the symmetry breaking down to the
Standard Model
\ba
{\cal H}_a &=&(1,3,\bar
3)_{(\hphantom{+}0,+1,-1)}
\ea
 According to~\cite{Glashow:1984gc}, two  Higgs replicas ($\alpha=1,2$)
 are required in order  to obtain both symmetry breaking to SM and
non-trivial quark mixing.

In the D-brane construction additional matter can arise from  strings with
both ends attached on the same brane stack. In particular, as we will see in
 the next sections, the following scalar fields are required in the present
 model in order to eliminate additional $Z'$ bosons
\ba
{\cal H}_{\cal L}&=&(1,3,1)_{(0,-2,0)}\label{HL}
\\
{\cal H}_{\cal R}&=&(1,1,3)_{(0,0,-2)}\label{HR}
\ea
These representations arise from strings having both their ends on
left and right brane-stacks respectively (see figure \ref{u333}).

Employing  the usual hypercharge embedding
\footnote{
The decomposition of the ${SU(3)}^3$ representations with respect to
$
SU(3)_C\times SU(3)_L\times SU(3)_R$ $\supseteqq$ $SU(3)_C\times
\left[SU(2)_L\times U(1)_{L'}\right]\,\times\,\left[U(1)_{R'}\times U(1)_{\Omega}\right]
$
are
\ba
(3,\bar 3,1)&\ra& (3,2;-1,0,0)+(3,1;2,0,0)\nonumber\\
(\bar 3,1,3)&\ra& (\bar 3,1;0,1,1)+(\bar 3,1;0,1,-1)+(\bar
3,1;0,-2,0)
\nonumber\\
(1,3,\bar 3)&\ra&(1,2;1,-1,1)+(1,1;-2,-1,1)+(1,2;1,-1,-1)+(1,1;-2,-1,-1)+1,2;1,2,0)+(1,1;-2,2,0)\nonumber
\ea
}
\ba
Y=-\frac{1}{6} X_{L'}+\frac 13 X_{R'}\label{HC}
\ea
(where $X_{L'}$ and $X_{R'}$ represent the $U(1)_{L'}$  and $U(1)_{R'}$ generators
respectively), the transformations of the fermion fields under $SU(3)_C\times SU(2)_L\times
U(1)_Y\times U(1)_{\Omega}$ are as follows (here we have suppressed the ${U(1)}_{C,L,R}$ indices):
\ba
{\cal Q}^{\hphantom{c}}&=&
q\left(3,2;\frac{1}{6},0\right)+g\left(3,1;-\frac{1}{3},0\right)\nonumber\\
{\cal Q}^c&=&
d^c\left(\bar3,1;\frac{1}{3},1\right)+u^c\left(\bar3,1;-\frac{2}{3},0\right)
+g^c\left(\bar3,1;\frac{1}{3},-1\right)\label{fermions}\\
{\cal L}^{\hphantom{c}} &=&
\ell^+\left(1,2;-\frac{1}{2},1\right)+\ell^-\left(1,2;-\frac{1}{2},-1\right)
+{\ell^c}\left(1,2;+\frac{1}{2},0\right)\nonumber\\&~&+
\nu^{c+}(1,1;0,1)+\nu^{c-}(1,1;0,-1)+e^c(1,1;1,0)
 \nonumber
\ea
Similarly, for the scalars we have
\ba
{\cal H}_a &=&(1,3,\bar
3)=h^{d+}_a\left(1,2;-\frac{1}{2},1\right)+h^{d-}_a\left(1,2,-\frac{1}{2},-1\right)
+{h^u}_a\left(1,2;\frac{1}{2},0\right)\nonumber\\
&~&+{e_H^c}_a(1,1;1,0)+{\nu_{H}^{c+}}_a(1,1;0,1)+{\nu_{H}^{c-}}_a(1,1;0,-1),\;\;\;
a=1,2,\dots\label{higgs}
\ea
 For the present work, we choose the following vevs
\ba
{\cal H}_1&\ra&\langle h^u_1  \rangle =u_1,\; \langle h^{d-}_{1}
\rangle =u_2,\;
                   \langle {{\nu^{c+}_{H,}}_1} \rangle =U,\label{H1}\\
{\cal H}_2&\ra&\langle h^u_2 \rangle =v_1,\; \langle {h^{d-}}_2
\rangle =v_2,\; \langle {h^{d+}}_2 \rangle =v_3, \langle
{\nu^{c-}_{H}}_2 \rangle =V_1,\; \langle {\nu^{c+}_{H}}_2 \rangle
=V_2.\label{H2}
\ea
 The vacuum expectation values $U,V_1,V_2$ are taken to be of the order of the
 $SU(3)_R$ breaking scale. Any of them, breaks the $SU(3)^3$ to an $SU(2)\times
 SU(2)\times U(1)$ residual symmetry,  whereas  two of them, namely $U$ and $V_1$,
 suffice to break the symmetry  down to $SU(2)\times U(1)$ Standard Model gauge group.
 The vevs $u_{1,2}, v_{1,2,3}$ related to the $SU(2)_L$
 Higgs doublets, should be taken of the order of the electroweak scale. In the sequel
 we shall assume for simplicity that  $V_i\sim U\sim M_R$.

One of the characteristics of string derived models is the appearance of anomalous
$U(1)$ symmetries. Unlike the heterotic string case where only one abelian factor
 is anomalous, in type I theory,  many anomalous abelian factors can be present and
 their cancellation is achieved through a generalized Green--Schwarz mechanism~\cite{kk}
 which utilizes the axion fields of the Ramond--Ramond
sector\cite{Ibanez:1999it,Lalak:1999bk}, providing masses to the
corresponding anomalous gauge bosons.

In the model under consideration, the mixed anomalies of the
non-abelian $SU(3)^3$ gauge part with the abelian $U(1)_{C,L,R}$
factors are proportional to ${\cal A}\sim{\rm Tr} Q_I T_J^2$ where
$T_J=\left\{{SU(3)}_C,{SU(3)}_L,{SU(3)}_R\right\}$ and
$Q_I=\left\{{U(1)}_C,{U(1)}_L,{U(1)}_R\right\}$ with
\ba
{\cal A}&=&\left(\begin{array}{ccc}
\hphantom{+}0&+1&-1\\
-1&\hphantom{+}0&\hphantom{+}1\\
+1&-1&\hphantom{+}0
\end{array}\right)
\ea
It is easy to see that there is only one anomaly-free $U(1)$
combination, namely
\ba
U(1)_{{\cal Z}'}&=& U(1)_C+U(1)_L+U(1)_R\label{aas}
\ea
 All  states represented from strings having their ends attached on two
 different brane stacks, i.e.  ${\cal Q},{\cal Q}^c, {\cal L}$ and ${\cal H}$
 in (\ref{fermions}) and (\ref{higgs}), have zero ``charge'' under ${\cal Z}'$.
  States represented by strings having both their
ends attached to the same brane stack, as is the case of ${\cal
H}_{\cal L}$ and ${\cal H}_{{\cal R}}$, are ``charged'' under
$U(1)_{{\cal Z}'}$. Under the standard hypercharge definition,
${\cal H}_{\cal L}$, ${\cal H}_{\cal R}$ are fractionally charged.
The standard hypercharge (\ref{HC}) is embedded in $SU(3)_L\times
SU(3)_R$, however,  it could also include the anomaly-free
$U(1)_{{\cal Z}'}$, so that $ Y'=Y+x {\cal Z}'$. This is possible
since, as explained above, the fermion and standard Higgs
multiplets carry zero $U(1)_{{\cal Z}'}$ charge, therefore, their
hypercharge is not affected. On the contrary, the fractionally
charged states ${\cal H}_{{\cal L},{\cal R}}$ will receive a
$U(1)_{{\cal Z}'}$-contribution in their hypercharge. Choosing  an
appropriate value for the coefficient $x$, the representations
${\cal H}_{\cal L}=(1,3,1)$ and ${\cal H}_{\cal R}=(1,1,3)$ obtain
integral charges like those of the standard model Higgs and lepton
fields. In particular, the embedding
\ba
Y'\;=\;Y+\frac 16{{\cal Z}'}\;\equiv\;-\frac{1}{6} X_L+
\frac 13 X_R+\frac 16{{\cal Z}'}\label{nHd}
\ea
leaves all the representations containing the SM spectrum
unchanged, while for the ${\cal H}_{\cal L, R}$ scalar fields it
yields~\footnote{The transormations of the fields ${\cal H}_{\cal L,R}$, under
$SU(3)_C\times \left[SU(2)_L\times U(1)_{L'}\right]\,\times\,$
$\left[U(1)_{R'}\times U(1)_{\Omega}\right] $ and $U(1)_{C,L,R}$,
are
\ba
{\cal H}_{\cal L}&=&\hat h_L(1,2,1;+1,0,0)_{(0,-2,0)}+\hat
\nu_{{\cal H}_{\cal L}}(1,1,1;-2,0,0)_{(0,-2,0)}\nonumber\\
{\cal H}_{\cal R}&=&\hat \nu_{{\cal H}_{\cal
R}}^+(1,1,1;-2,-1,+1)_{(0,0,-2)}+\hat \nu_{{\cal H}_{\cal
R}}^-(1,1,1;-2,-1,-1)_{(0,0,-2)}+\hat e_{{\cal H}_{\cal
R}}^0(1,1,1;-2,2,0)_{(0,0,-2)} \nonumber
\ea }
\ba
{\cal H}_{\cal L} &=&(1,3,1)=\hat
h_L^{+}\left(1,2;-\frac{1}{2},0\right)+\hat\nu_{{\cal H}_{\cal
L}}\left(1,1;1,0\right)
\\
{\cal H}_{\cal R} &=&(1,1,3)= {\hat
e_H^c}(1,1;1,0)+{\hat\nu_{{\cal H}_{\cal
R}}^{c+}}(1,1;0,1)+\hat\nu_{{\cal H}_{\cal R}}^{c-}(1,1;0,-1)
\ea
where, an in the case of the representations in (\ref{fermions}),
the transformation properties and the quantum numbers of ${\cal
H}_{\cal L,R}$ are written here with respect to the symmetry
$SU(3)_C\times SU(2)_L \times U(1)_Y\times U(1)_{\Omega}$. Thus,
under (\ref{nHd}) the multiplet ${\cal H}_{\cal L}$ contains a
standard Higgs doublet $\hat h_L$ and a neutral singlet
$\hat\nu_{{\cal H}_{\cal L}}$. The ${\cal H}_{\cal R}$
representation is  decomposed into a charged singlet $\hat
e_{{\cal H}_{\cal R}}$ and the two neutral components
$\hat\nu^+_{{\cal H}_{\cal R}}$, $\hat\nu^-_{{\cal H}_{\cal R}}$
which will play a crucial role to the formation of heavy mass
terms for the additional lepton doublets and the breaking of the
extra $U(1)_{{\cal Z}'}$.  Indeed,  since ${\cal H}_{1,2}$  Higgs
fields do not carry any charge under $U(1)_{{\cal Z}'}$, the
latter remains unbroken. Thus, to break this remnant abelian
factor, we need to assume non-zero vevs for the ${\cal H}_{\cal
L}$ and/or ${\cal H}_{\cal R}$ field.

\subsection{The string and the weak angle}

In a D-brane realization of the proposed model, the three $U(3)$
gauge factors  originate from 3-brane stacks that span different
directions of the higher dimensional space. As a consequence, the
corresponding  gauge couplings $\alpha_{C,L,R}$ are not
necessarily equal at the string scale $M_S$. This is a general
properly of Type I string constructions where the volume enters
the relation between the string and the gauge couplings, in
contrast to the Heterotic string case. However, in certain
constructions, at least two D-brane stacks can be superposed and
the associated couplings are equal\cite{Antoniadis:2002en}. In
this scenario the low energy data together with the gauge coupling
running can be used  to determine the string scale $M_S$
\cite{Gioutsos:2005uw}.
In this context, we examine three different cases (i)
$\alpha_L=\alpha_R\equiv a$, (ii) $\alpha_C=\alpha_L\equiv a$ and
(iii) $\alpha_C=\alpha_R\equiv a$ at $M_S$ which correspond to
superposing the left with the right,
 the color with the left and the color with the right $U(3)$ brane stacks.

The reduction of the ${SU(3)}^3\times{U(1)^3}$ to the SM  is in
general associated with three different scales corresponding to
the  the $SU(3)_R$,  $SU(3)_L$ and $U(1)_{{\cal Z}'}$ symmetry
breaking. We will assume here for simplicity that the
$SU(3)_{L,R}$ and $U(1)_{{\cal Z}'}$ symmetries break
simultaneously at a common scale $M_R$, hence  the model is
characterized only by two large scales, the string/brane scale
$M_S$, and the  scale $M_R$.~\footnote{For a detailed analysis
see~\cite{future}.} Clearly, the $M_R$ scale cannot be higher than
$M_S$, i.e., $M_R\le M_S$, and the equality holds if the
$SU(3)_R\times SU(3)_L$ symmetry breaks directly at $M_S$. In
order to determine the range of $M_S,M_R$, we use as inputs the
low energy data for $\alpha_3,\alpha_{em}$ and $\sin^2\theta_W$
and perform a one-loop renormalization group analysis. Taking into
account the $U(1)$ factor normalizations the hypercharge embedding
(\ref{nHd},\ref{aas}), implies for $\mu\ge M_R$
\ba
\frac{1}{\alpha_Y}&=&\frac 12\frac{1}{\alpha_L}+\frac
32\frac{1}{\alpha_R}+\frac 16\frac{1}{\alpha_C}.\label{fHg}
\ea
As a consequence of (\ref{fHg}) the weak angle at the string scale is given by
\ba
\sin^2\theta_W&=&\frac{6}{9\left(1+\frac{\alpha_L}{\alpha_R}\right)+\frac{\alpha_L}{\alpha_C}}
\ea
Thus, for equality of all gauge couplings
$\alpha_C=\alpha_L=\alpha_R$ at $M_S$, we obtain
$\sin^2\theta_W=\frac{6}{19}$.

\begin{figure}[!t]
\centering
\includegraphics[scale=.9]{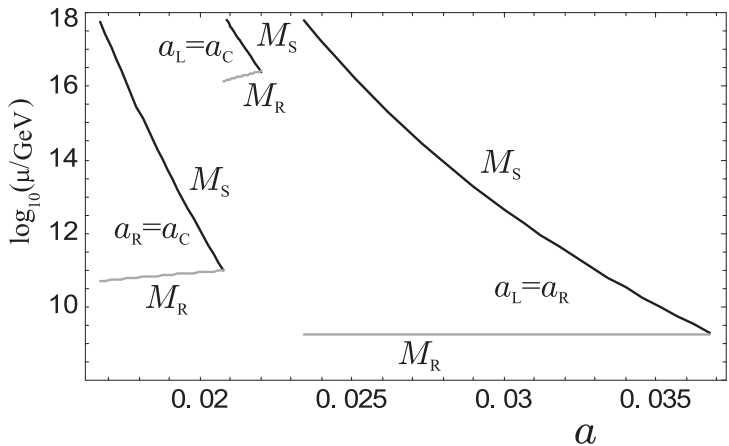}
\caption{The string scale $M_S$, and $SU(3)_R$ breaking scale $M_R$
 as functions of the common coupling $a$ for (i) $\alpha_L=\alpha_R=a$,
(ii)  $\alpha_L=\alpha_C=a$ and (iii) $\alpha_C=\alpha_R=a$. In all cases, we let
$M_S$ is truncated at  $ 10^{18}$GeV.
In case (i) the $M_R$ scale is constant
$M_R\approx 1.7\times 10^{9}GeV$. In the remaining two cases, we find that
 $M_R$ lowers as $M_S$ attains higher values. }
 \label{msmr}
\end{figure}

The  one loop renormalization group equations are: $\alpha_i^{-1}(\mu)=\alpha_i^{-1}(M_Z)-\frac{b_i}{2\,\pi}
\ln\frac{\mu}{M_Z}$ where $i=2,3,Y$ for $M_Z\le\mu\le M_R$ and
$\alpha_i^{-1}(\mu)=\alpha_i^{-1}(M_R)-\frac{b_j'}{2\,\pi}
\ln\frac{\mu}{M_Z}$ where $j=C,L,R,Z'$ for $M_R\le\mu\le M_S$. The two Higgs SM beta functions are:
$b_3=-7, b_2=-3, b_Y=7$ and the ${SU(3)}^3$ beta functions are: $b_{C}'=-5$,
$b_{L}'=b_{R}'=-\frac{59}{12}+\frac{n_{\hat H}}{4}\,$,
where  $n_{\hat H}$
the number of the Higgs fields ${\cal H}_{{\cal L},{\cal R}}$ which in our case is taken to be $n_{\hat H}=2$.
Solving the RGEs for the three cases mentioned above we obtain $M_R$ and $M_S$ as a function
of the common coupling $a$. The results are presented in Figure \ref{msmr}.
The curves extend from the point $M_S=M_R$ to  the Planck
scale.

 The case in the right part of the graph corresponds to $\alpha_L=\alpha_R=a$.
 We observe that in this case, the $M_R$ scale remains constant
  $M_R\sim 1.7\times 10^9$\,GeV, i.e., it is independent of  the common gauge
  coupling $a$.  The second case (in the middle of the graph)  corresponds to
  the case $\alpha_L=\alpha_C=\alpha$. The
identification of $M_S,M_R$ scales occurs at the unification point
$M_S=M_R\approx 2.3\times 10^{16}$GeV. Finally, for
$\alpha_R=\alpha_C$, we obtain $M_R=M_S\approx
2.3\times10^{11}$GeV. The bounds on the $M_S,M_R$ scales for the
three cases under consideration are summarized in Table
\ref{ytab3a}.
\begin{table}[!h]
\centering
\begin{tabular}{|c|c|l|}
\hline model&$M_R/GeV$ &${M_S}/{GeV}$\\
\hline
 $a_L=a_R$&$
 1.7\times 10^{9}$&$
 >1.7\times 10^{9}$\\
\hline $a_L=a_C$&$
 <2.3\times 10^{16}$&$
>2.3\times 10^{16}$\\
 \hline
$a_C=a_R$&$
 <2.3\times 10^{11}$&$
>2.3\times 10^{11}$\\
 \hline
\end{tabular}
\caption{\label{ytab3a} Upper and lower bounds for $SU(3)_R$
breaking scale ($M_R$)  and the corresponding string scale ($M_S$)
for the three cases $a_L=a_C$, $a_R=a_C$ and $a_L=a_R$. }
\end{table}

We finally consider the case of unification of all couplings
$\alpha_C=\alpha_R=\alpha_L$ at $M_S$. Using the renormalization
group equations and the fact that the spectrum of the model
implies $b_L=b_R$ we find that the $M_R$ scale does not depend on
$b_{L,R,C}$ beta functions and can be expressed only in terms of
the low energy parameters as follows
\ba
M_R&=&M_Z\times
\exp\left[\frac{6/a_Y-12/a_2-1/a_3}{6b_Y-12b_2-b_3}\right]\approx
1.7\times 10^{9} GeV
\ea
Thus, $M_R$, at least in the one loop approximation, is
independent  of  the physics at the string scale. On the contrary,
$M_S$, as expected, strongly depends on the $SU(3)^3$ beta
functions and turns out to be rather high for the minimal content.
Demanding $M_S\sim 4\times 10^{17}$ GeV
 implies that the beta-functions $b_{L,R}$ should be at least $b_{L,R}\ge -\frac{3}{2}$. Such
values for the $b_{L,R}$ beta functions are  obtained for a large number of Higgs fields
 and other matter multiplets which are usually present in a string spectrum.

\section{Yukawa couplings and fermion masses}

We turn now to the fermion mass problem. In contrast to the
${SU(3)}^3$ model \cite{Glashow:1984gc,Rizov:1981dp} in our
$U(3)^3$ construction the tree level Yukawa potential consists of
a single fermion mass term. This is due to the existence of the
additional $U(1)_{C,L,R}$ symmetries which eliminate other
possible mass terms.
 Indeed,  only the coupling
\ba
\lambda_{Q,a}^{ij}{\cal Q}_i\,{\cal Q}^c_j\,H_a,\;\; i,j=1,2,3\ ,\ \alpha=1,2\label{mm}
\ea
is allowed at tree-level,  providing up and down quark masses as
well masses for the extra triplets. More precisely, for the Higgs breaking pattern
described in (\ref{H1}),(\ref{H2}) the following mass terms arise
from (\ref{mm}) for the up quarks
\ba
m_{uu^c}^{ij}\;u_i\,u_j^c&=&(\lambda_{Q,1}^{ij}u_1+\lambda_{Q,2}^{ij}v_1)\,u_iu_j^c\label{uqm}
\ea
For the down-type quarks $d_i,d_j^c, g_i, g_j^c$, we obtain a
$6\times 6$ mass matrix in flavor space, of the form
\ba
m_{dg}&=&
\bordermatrix{
&d&g\cr
d^c&\lambda_{Q,1}^{ij}\,u_2+\lambda_{Q,2}^{ij}\,v_2&\lambda_{Q,2}^{ij}\,V_1\cr
g^c&\lambda_{Q,2}^{ij}\,v_3&\lambda_{Q,1}^{ij}\,U+\lambda_{Q,2}^{ij}\,V_2}
\label{mdg}
\ea
As seen from (\ref{mdg}) the $m_{dd^c}$ and the
$m_{dg^c}$ 3$\times$3 sub-matrices are of the order of the
electroweak scale, whilst $M_{dg^c}, M_{gg^c}$ are of the order
$M_R$.  The diagonalization of the full non-symmetric mass matrix
(\ref{mdg}) results to light   masses for the
down quarks and masses of the order  $M_R$ for the extra states.
 Few comments are in order.
Realistic quark mixing~\cite{Glashow:1984gc} implies the necessity of
two replicas of Higgs representations. Indeed,
checking the structure of the mass terms in (\ref{mm}) and (\ref{mdg}),
 we observe that if a single Higgs field is present,
the up and down quark mass matrices are proportional
$\lambda_{Q,1}^{ij}u_1\propto \lambda_{Q,1}^{ij}u_2$ resulting to
absence of  KM mixing~\footnote{String/brane models can provide
a different solution to this problem, as the flavor matrix may depend
on geometric quantities~\cite{Cremades:2003qj}.}.
 Note further that we may set $v_3=0$ which leads to
$m_{dg^c}=0$. Moreover, assuming $V_2=0$, there is no substantial
change on the matrix. If both $v_3=0$ and $V_1=0$, then $d$ and
$g$ quarks decouple completely.

We turn now our attention to the lepton mass matrices.  Due to the
three extra $U(1)$ factors, tree-level lepton masses are not
allowed. In particular, the $U(1)_{C,L,R}$ `charges' do not allow
for a coupling of the form ${\cal L}\,{\cal L}\,{\cal H}_i$. The
lowest order allowed leptonic mass term arises at fourth order
\ba
\frac{f_{ij}^{ab}}{M}\, {\cal H}_a^\dagger \,{\cal
H}_b^\dagger\,{\cal L}_i\,{\cal L}_j
\ea
This term provides electroweak scale masses for the charged
leptons as well as Dirac and  Majorana neutrino masses.
For the charged leptons we have, considering  all possible combinations
of the two Higgs doublets, i.e., ${\cal H}_1{\cal H}_1, {\cal H}_1{\cal H}_2$
and ${\cal H}_2{\cal H}_2$, we obtain the mass terms
\ba
\frac{f_{ij}^{ab}}{M}\,{\cal H}_a^\dagger \,{\cal
H}_b^\dagger\,{\cal L}_i\,{\cal L}_j&\ra &
\frac{f_{ij}^{ab}}{M}\left(\langle h_{a}^{u*}\rangle\,\langle
\nu_{H\,b}^{c+*}\rangle\,\ell^+_i e^c_j + \langle
h^{u*}\rangle\,\langle \nu_{H\,b}^{c-*}\rangle\,\ell_i^-
e^c_j\right)\nonumber
\\&\ra&  \left(\alpha_{ij}\,\ell^+_i +
\beta_{ij}\;\ell^-_i \right)\,e^c_j\label{lmt}
\ea
where
\ba
\alpha_{ij}&=&
\rho(f_{ij}^{11}u_1+f_{ij}^{21}v_1)+\sigma (f_{ij}^{22}v_1+f_{ij}^{12}u_1)\nonumber\\
\beta_{ij}&=&\xi\left(f_{ij}^{22}v_1+f_{ij}^{21}u_1\right)
\ea
and $\rho=\frac{U}{M}, \sigma=\frac{V_2}M, \xi=\frac{V_1}M$.

 Thus, up to this point, we find that the linear
combination $\alpha_{ij}\,\ell^+_i +
\beta_{ij}\,\ell^-_i$ defines the light left-handed lepton doublet
$\ell_i$  which couples to the right handed electron $e^c_j$.
Clearly, there is an orthogonal to the above linear combination
which, together with the anti-lepton doublet $\ell^c$ remain
massless at this level. It would be necessary to obtain a heavy
mass for this remaining pair of doublets. Surprisingly, the
additional Higgs fields ${\cal H}_{\cal L,R}$ introduced in order
to break $U(1)_{{\cal Z}'}$ residual symmetry  generate heavy
masses for these extra doublets. The  Yukawa couplings involving
the Higgs fields ${\cal H}_{\cal L}$, ${\cal H}_{\cal R}$ are of
the form
\ba
\frac{\zeta_{ij}}{M}\,{\cal H}_{\cal L} \,{\cal H}_{\cal R}^\dagger\,{\cal L}_i\,{\cal
L}_j&\ra &\langle \hat h_L^+\rangle \left(\hat\alpha_{ij}
\ell^+_i+\hat\beta_{ij}\ell^-_i\right)\,e^c_j
+\langle \hat \nu_{{\cal H}_{\cal L}}\rangle \left(\hat\alpha_{ij}
\ell^+_i+\hat\beta_{ij}\ell^-_i\right)\,\ell^c_j
\ea
with $\hat\alpha_{ij}=\zeta_{ij} \frac{\langle\hat
\nu_{{\cal H}_{\cal R}}^{c+*}\rangle}M, \hat\beta_{ij}=\zeta_{ij}\frac{\langle
\hat\nu_{{\cal H}_{\cal R}}^{c-*}\rangle}M$.

In the above coupling, without loss of generality, we may assume
$\langle \hat{h}_{L}^+\rangle =0$. The remaining vevs $\langle \hat
\nu_{{\cal H}_{\cal L}}\rangle,\langle \hat \nu_{{\cal H}_{\cal R}}^{c+*}\rangle, \langle
\hat\nu_{{\cal H}_{\cal R}}^{c-*}\rangle$ should be taken of the order
$M_R$~\footnote{These vevs have far reaching consequences to the
symmetry breaking, since now one dispenses with the use of a
second Higgs ${\cal H}=(1,3,\bar 3)$ to break $U(3)^3$ down to
SM\cite{future}.} so that the heavy linear combination for leptons
$\hat \ell_i=\hat\alpha_{ij} \ell^+_i+\hat\beta_{ij}\ell^-_i$
couples to the antilepton doublet $\ell^c$ to form a massive state
of the order $M_R$.

We now turn our attention to the neutral lepton states.
 Suppressing the order one Yukawa
coefficients ($f_{ij}^{ab}$),  the neutrino mass matrix
 in the
basis $\ell^+,\ell^-,\nu^{c+},\nu^{c-},\ell^c$ is
\ba
M_{\nu}&\sim&\frac{1}{M_S}
\left(\begin{array}{ccccc}
m_W^2&m_W^2&m_W\,M_R&M_R^2&M_R^2\\
m_W^2&m_W\,M_R&m_W\,M_R&m_W\,M_R&M_R^2\\
m_W\,M_R&m_W\,M_R&M_R^2&M_R^2&0\\
m_W\,M_R&m_W\,M_R&M_R^2&M_R^2&0\\
M_R^2&M_R^2&0&0&0\\
\end{array}
\right)\nonumber
\ea
where we have assumed for simplicity  common vevs $u_i=v_i=M_W$,
$U=V_j=M_R$, where $i,j=1,2$. This is a see-saw type mass matrix. Three light
neutrino species receive see-saw masses of the order $m_W^2/M_S$
while the remaining states receive heavy masses of the order
${M_R^2}/{M_S}$. To obtain a light neutrino spectrum at the range
of $eV$, the scale $M_S$ should be of the order $M_S\sim 10^{13-15}$GeV.
Interestingly, this is in accordance with the findings of
the RGE analysis in section 3. In particular, $M_S$ is found within the bounds
of the cases $\alpha_L=\alpha_R$ and $\alpha_C=\alpha_R$ shown
in figure 2. It is further compatible with the effective gravity
scale in theories with large extra  dimensions  obtained in the
context of Type I string models~\cite{Blumenhagen:2000wh}.

\section{Conclusions}

Inspired by D-brane scenarios, in this work, we analyzed a
non-supersymmetric $U(3)_C\times U(3)_L\times U(3)_R$ model which
is equivalent to the standard $SU(3)_C\times SU(3)_L\times
SU(3)_R$ ``trinification'' gauge group  supplemented by three
$U(1)_{C,L,R}$ factors. The three additional abelian factors have
mixed anomalies with $SU(3)_{C,L,R}$ generators. These anomalies
can be cancelled in the context of a D-brane derived model
 by a generalized Green-Schwartz mechanism.
The single anomaly free combination $U(1)_{{\cal Z}'}=U(1)_C+
U(1)_L+U(1)_R$ contributes to the hypercharge.

The Standard Model fermions, represented by strings attached to
two different brane-stacks,  belong to $(3,\bar 3,1)+(\bar
3,1,3)+(1,3,\bar 3)$ representations as in the case of the
``trinification'' model. Two Higgs fields in $(1,3,\bar 3)$ (in the
same representation as the lepton fields) and two
in $(1,3,1)$ and $(1,1,3)$ representation can acquire vevs that completely break the
gauge symmetry to the SM. They also provide masses to
all additional matter fields.
 The model is characterized by  a natural quark-lepton hierarchy since quark masses
are obtained from tree-level couplings, while, due to the extra
$U(1)$ symmetries, charged leptons are allowed to receive masses
from fourth order Yukawa terms.
 Light Majorana neutrino masses are obtained
 through a see-saw type mechanism operative at the scale  $M_R$ which turns out to be
 $M_R\ge10^9$GeV.

{\bf Acknowledgements}. {\it This research was funded by the program
`PYTHAGORAS' (no.\ 1705 project 23) of the Operational
 Program for Education and Initial Vocational Training of the Hellenic
 Ministry of Education under the 3rd Community Support Framework and
 the European Social Fund. The authors would like to acknowledge kind hospitality
 of CERN, where part of this work has been completed.}

\end{document}